\documentclass[reprint,secnumarabic, amssymb, amsmath, aps,prb]{revtex4-1}

\usepackage{docs,graphicx,bm}
\usepackage[colorlinks=true,linkcolor=blue]{hyperref}%
\expandafter\ifx\csname package@font\endcsname\relax\else
\expandafter\expandafter
\expandafter\usepackage
\expandafter\expandafter
\expandafter{\csname package@font\endcsname}%
\fi
\begin{document}

\title{Amplified-reflection plasmon instabilities in grating-gate plasmonic crystals}

\author{Aleksandr S. Petrov}%
\email{aleksandr.petrov@phystech.edu}
\affiliation{Laboratory of 2D Materials' Optoelectronics, Moscow Institute of Physics and Technology, Dolgoprudny 141700,	Russia}%

\author{Dmitry Svintsov}%
\affiliation{Laboratory of 2D Materials' Optoelectronics, Moscow Institute of Physics and Technology, Dolgoprudny 141700,	Russia}%

\author{Victor Ryzhii}%
\affiliation{Institute of Ultra High Frequency Semiconductor Electronics RAS,
Moscow 117105, Russia}
\affiliation{Research Institute of Electrical Communication, Tohoku University, Sendai 980-8577\,Japan}%

\author{Michael S. Shur}%
\affiliation{Department of Electrical, Electronic, and System Engineering and Department of Physics, Applied Physics, and Astronomy, Rensselaer Polytechnic Institute, Troy, New York 12180, USA
}

\date{October 2016}%

\begin{abstract}
    We identify a possible mechanism of the plasmon instabilities in periodically gated two-dimensional electron systems with a modulated electron density (plasmonic crystals) under direct current. The instability occurs due to the amplified reflection of the small density perturbations from the gated/ungated boundaries under the proper phase matching conditions between the crystal unit cells. Based on the transfer-matrix formalism, we derive the generic dispersion equation for the travelling plasmons in these structures. Its solution in the hydrodynamic limit shows that the threshold drift velocity for the instability can be tuned below the plasmon phase and carrier saturation velocities, and the plasmon increment can exceed the collisional damping rate typical to III-V semiconductors at 77\,K and graphene at room temperature. 
    \end{abstract}

\maketitle

\section{Introduction}

The emission of terahertz radiation from two-dimensional electron systems (2DES) under direct current flow has been observed in a large number of experiments, starting from the pioneering work of Tsui, Gornik and Logan~\cite{tsui1980firEmission}. At the current stage of technology, the emission from these structures sustains up to the room temperature~\cite{dyakonova2006room}, its frequency is voltage-tunable from $0.5$ to 2 THz~\cite{ElFatimy_2010_algan}, while the linewidth can be as narrow as $\sim 40$ GHz~\cite{otsuji2013emission}. It is commonly accepted that the radiation appears as a result of plasmon excitation~\cite{tsui1980firEmission,matov1997generation,mikhailov1998plasma,dyakonov1993shallow} in 2DES and the subsequent coupling of plasmon to the free-space radiation upon interaction with single~\cite{dyakonov1993shallow} or multiple~\cite{krasheninnikov1985radiative} metal gates. The periodically gated 2DES typically demonstrate emission of higher power and narrower linewidth~\cite{Otsuji2007SSEEmission,Otsuji2008JPCMemission,otsuji2013emission} compared to the plasmonic transistors with a single gate. Despite these experimental advances, there exists no accepted theory on the mechanism of plasmon self-excitation in grating-gated plasmonic nanostructures.

Early works have suggested the excitation of plasmons by hot electrons~\cite{Chaplik1985absorptionemission}. However, in the latest experiments~\cite{otsuji2013emission} the emission sets on in a threshold-like manner, which signifies the occurrence of plasma instability. In the simplest case of dc electron flow in 2DES parallel to the conducting gate, the dissipative instabilities~\cite{krasheninnikov1980instabilities} can develop at drift velocity equal to the plasmon velocity. A similar estimate of threshold velocity was obtained for amplified transmission of radiation through periodically-gated 2DES with uniform density~\cite{matov1997generation,mikhailov1998plasma}.  Such high velocity can be hardly achieved in experiment, particularly, due to the choking of electron flow~\cite{dyakonov1995choking}.

The onset of terahertz emission in grating-gated 2DES at low longitudinal electric field ($\sim 1$ kV/cm in \cite{Otsuji2008JPCMemission}) motivated the search of low-threshold plasmon instabilities. In Refs.~\cite{ryzhii2005transit,ryzhii2008mechanism,Koseki2016Giant} it was supposed the latter can emerge due to the transit time effects in the high-field domains of 2DES. However, the transit time effects generally require the deviations from linear relation between current and electric field, e.g. due to velocity saturation. The voltage drop across each cell of experimentally relevant grating-gated 2DES~\cite{Otsuji2007SSEEmission} is less than 20 mV, and the transit time effects can be suppressed at such low voltages.

\begin{figure}[t]
\includegraphics[width=0.9\linewidth]{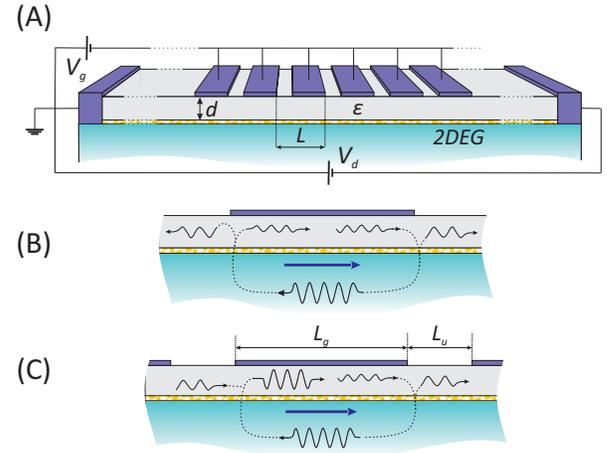}
\caption{\label{Fig1} 
(A) Schematic view of a field-effect transistor with two-dimensional conducting channel and periodic grating-gate structure. (B, C) Schematic view of plasmon reflections from gated/ungated boundaries under an isolated gate in an infinite 2DEG (B) and in the multigate structure (C).}
\end{figure}

A new class of plasma instabilities based on amplified plasmon reflection in bounded 2DES was put forward by Dyakonov and Shur~\cite{dyakonov1993shallow}. Their threshold velocity is limited only by the carrier scattering by impurities or phonons, and can be made very low in sufficiently clean systems. At the same time, the Dyakonov-Shur (DS) instability relies on essentially asymmetric boundary conditions at the 2DES contacts: the impedance at the drain should be greater than at the source~\cite{Cheremisin1999}. Such asymmetry is not present in a weakly biased {\it isolated} gated cell in an infinite 2DES (Fig.~1B).

In the present paper, we theoretically show that the reflection-type plasma instabilities can develop in {\it periodically} gated 2DES (plasmonic crystals) with a modulated electron density. We find that the travelling waves with quasi-momentum not at the edge of the plasmonic Brillouin zone are generally more unstable than purely periodic waves considered in Ref.~\cite{mikhailov1998plasma}. We show that the drift velocity required for the emergence of unstable modes can be well below the plasmon phase velocity. This contrasts to the case of 'plasmonic boom' instabilities in all-gated plasmonic crystals with a varying carrier density~\cite{kachorovskii2012current} or of a varying width~\cite{Aizin2016plasmonicboom} that occur at 'superplasmonic' drift velocities. Remarkably, the proposed mechanism of instability requires neither transit-time nor velocity saturation effects~\cite{ryzhii2005transit,ryzhii2008mechanism}. It can be thus responsible for the plasmon instabilities and THz emission in grating-gate structures with graphene and GaN channels, where the critical field and saturation velocity are very large~\cite{meric2008current_saturation}.

Our mechanism of instability can be understood as follows. In a single gated cell of plasmonic crystal, the downstream plasmon undergoes an amplified Dyakonov-Shur reflection from the gated/ungated boundary~\cite{Dyakonov2008boundaryinstability,Sydoruk2012mirrors}. In an isolated gated cell, the reflected upstream wave would be attenuated upon reflection from the opposite boundary. But under the proper phase matching conditions in multigate structures, the fraction of the plasmon energy from the previous cell can compensate the reflection loss of the upstream wave (Fig.~\ref{Fig1}C). This periodic amplified reflection results in the net instability.

In Sec.~II, we derive the generic dispersion equation for the plasmons in a periodically gated 2DES in the presence of the electron drift and discuss the stability of the solutions. In Sec.~III, we find the eigenfrequencies and instability increments of the plasmonic modes in the hydrodynamic limit. Section IV discusses the possible experimental manifestations of this instability and further extensions of the model. 

\section{Plasmon dispersion and conditions of instability}

To provide a quantitative picture of the instability, we derive the dispersion equation for the travelling waves with a non-zero Bloch phase $\theta = q L$, where $q$ is the quasi-momentum and $L$ is the length of plasmonic crystal cell. According to the microscopic studies of wave reflection at the gated/ungated boundary\cite{Sydoruk2015reflection}, the net amplitude of wave can be approximated as a sum of the 'fast' downstream and 'slow' upstream plasmons both in the gated and ungated sections. This so-called quasi-optical approximation provides sufficient accuracy for long ungated sections and/or high frequencies~\cite{Shur2014gratingdispersion}. We denote the plasmon wave vectors as $k^{\rm g,u}_\pm$, where the plus sign stands for the downstream waves, and minus sign -- for the upstream waves, the superscripts 'g' and 'u' denote the gated and ungated sections, respectively. The amplitudes of electric potential in these waves are denoted as $\delta\varphi^{\rm g,u}_\pm$.

Let us compose the vectors $\delta {\boldsymbol \varphi}^{\rm g,u} = \{ \delta\varphi^{\rm g,u}_+,\delta\varphi^{\rm g,u}_-\}^{\rm T}$. In the neighboring cells of the crystal, they can differ only by the factor $e^{i \theta}$. On the other hand, they are related via the transfer matrix of the unit cell ${\hat T}$, $\delta{\boldsymbol \varphi}_{N+1} = \hat{T} \delta{\boldsymbol \varphi}_{N}$. This leads us to the general dispersion equation~\cite{pendry1994photonic}
\begin{equation}
    \mathrm{det}\left(\hat{T}-\hat{I}e^{i\theta}\right)=0,
    \label{eq-generalDispRel}
\end{equation}
where $\hat I$ is the identity matrix. The transfer matrix of the unit cell is represented as the product of transfer matrix characterizing the ungated section ${\hat T}_{\rm u}$, the matrix of the wave reflection and transmission at the ungated/gated boundary ${\hat T}_{\rm b}$, the transfer matrix of the gated part ${\hat T}_{\rm g}$, and, finally, the $T$-matrix of another boundary:
\begin{equation}
    {\hat T} = {\hat T}_{\rm u} \cdot {\hat T}_{\rm b} \cdot {\hat T}_{\rm g} \cdot {\hat T}^{-1}_{\rm b}.
\end{equation}

The transfer matrices of free wave propagation ${\hat T}_{\rm g, u}$ have the diagonal form
\begin{equation}
    \hat{T}_{\rm g,u} = \begin{pmatrix} e^{ik^{\rm g,u}_+ L_{\rm g,u}} & 0 \\ 0 & e^{ik^{\rm g,u}_- L_{\rm g,u}}\end{pmatrix},
\end{equation}
where $L_{\rm g}$ and $L_{\rm u}$ are the lengths of the gated and ungated regions, respectively. Keeping in mind the flow-induced non-reciprocity, we can present the $T$-matrix describing the boundary as~\cite{Sydoruk2015THzplasmonscoupled}
\begin{equation}
    \hat{T}_{b} = \frac{1}{t_+} \begin{pmatrix} 1 & -r_- \\ r_+ & t_-t_+ - r_- r_+ \end{pmatrix}.
\end{equation}
where $r_{+(-)}$ and $t_{+(-)}$ are the reflection and transmission coefficients of the waves incident from the ungated (gated) region. These reflection and transmission coefficients are essentially different due to the presence of electron flow.

Known all elements of $T$-matrices, one readily obtains the dispersion equation for plasmons in drifting 2DEG with grating gate:
\begin{widetext}
\begin{multline}
\label{eq-Dispersion}
        \cos{\left(\theta+\frac{k^{\rm g}_+ + k^{\rm g}_-}{2}L_{\rm g}+\frac{k^{\rm u}_+ + k^{\rm u}_-}{2}L_{\rm u}\right)}= \\
        =\cos{\left(\frac{k^{\rm g}_+ - k^{\rm g}_-}{2}L_{\rm g}\right)} \cos{\left(\frac{k^{\rm u}_+ - k^{\rm u}_-}{2}L_{\rm u}\right)} -  Z\sin{\left(\frac{k^{\rm g}_+ - k^{\rm g}_-}{2}L_{\rm g}\right)} \sin{\left(\frac{k^{\rm u}_+ - k^{\rm u}_-}{2} L_{\rm u} \right)},
\end{multline}
\end{widetext}
where 
\begin{equation}
\label{ModDepth}
Z=1-2\,r_+ r_-/(t_+ t_-)    
\end{equation}
is the 'modulation depth' factor. Equation (\ref{eq-Dispersion}) is generic and its functional form does not depend on the transport properties in gated and ungated sections and at the boundaries. In the absence of a drift, it is similar to the plasmon dispersion in a fully gated 2DES with a modulated density\cite{kachorovskii2012current}, or to the photon dispersion in one-dimensional photonic crystals~\cite{Bendickson1996photonic1d} and electron dispersion in the Kronig-Penney potential. All the information about 'bulk' carrier transport in Eq.~(\ref{eq-Dispersion}) is contained in the plasmon wave vectors $k^{\rm g}$ and $k^{\rm u}$, while all the information about boundary transport is enclosed in the coefficients $r$ and $t$.

The instability conditions for the waves with the dispersion relation given by (\ref{eq-Dispersion}) can be derived in a very general form. As the system approaches the instability threshold with an increasing drift velocity, the stable anticrossing of the plasmon bands transforms into an unstable one (Fig.~2)~\cite{landau1993course}. Upon increasing the flow velocity, the plasmon band gap gradually decreases, and the shrinkage of the gap indicates the onset of instability. For a real-valued coefficient $Z$, this shrinkage occurs when the absolute value of the right-hand side of Eq.~(\ref{eq-Dispersion}) equals unity, while its frequency derivative equals zero. If the frequency dependence of the modulation depth is weak (this assumption will be justified in the next section), the instability conditions can be expressed in a concise form 
\begin{gather}
    Z = 1, \label{eq-Z}\\
   \frac{ k^{\rm g}_+ - k^{\rm g}_-}{2}L_{\rm g} + \frac{k^{\rm u}_+ - k^{\rm u}_-}{2}L_{\rm u} = \pi m,\; m\in Z. 
    \label{eq-piM}
\end{gather}
Equations~(\ref{eq-Z}) and (\ref{eq-piM}) allow one to determine simultaneously the critical flow velocity, the frequency and quasi-wave vector at which the instabilities occur.

\section{Analysis of the dispersion relation}

The usefulness of Eq.~(\ref{eq-Dispersion}) stems from the fact that it can be applied to a wide class of two-dimensional systems, once their plasmon dispersion relations $k(\omega)$ in the presence of drift are known. The calculation of the reflection and transmission coefficients governing the value of $Z$ can be also done by various methods differing in complexity and accuracy~\cite{sydoruk2012distributed,Sydoruk2015reflection}.

For the numerical estimates of the critical velocity and the magnitude of wave increment, we derive the frequency dependencies of the wave vectors $k^{\rm g,u}_{\pm}$ within the hydrodynamic model. The latter is justified when carrier-carrier collision frequency exceeds the plasma frequency and the frequency of the carrier collisions with impurities and phonons. Both the theoretical estimates~\cite{dyakonov1993shallow,Svintsov2012hydrodynamic} and experiments~\cite{DeJong-HD-Flow,Rudin2014experiment} support the applicability of the hydrodynamic model up to the terahertz frequencies in III-V 2DEGs and graphene \cite{Bandurin-HDgraphene}. The simultaneous solution of Poisson, Euler and continuity equations leads us to \begin{gather}
    k_{\pm}^{\rm g} =\frac{\omega}{u_{\rm g}\pm s} \label{eq-kg}\\
    k_{\pm}^{\rm u} = \frac{\omega u_{\rm u}\pm a\mp\sqrt{a^2\pm 2a\omega u_{\rm u}}}{u_{\rm u}^2},
    \label{eq-kug}
\end{gather} 
where $\omega$ is the plasma wave frequency, $u_{\rm g,\rm u}$ are the carrier drift velocities, $s=\sqrt{e V_g/m^*}$ is the plasma wave velocity in the absence of drift, $a=\pi e^2 n_u/(\varepsilon m^*)$, $n_u$ is the carrier density in the ungated region, $V_g$ is the gate-to-channel bias, $m^*$ is the electron effective mass, which we take to be $0.067m_0$, $\varepsilon$ is the gate dielectric constant.

The determination of the reflection and transmission coefficients requires an imposition of the boundary conditions for the electric potential, drift velocity and carrier density at the gated/ungated interface. If the electron transport obeys the hydrodynamic equations at the transient regions as well, the boundary conditions for the determination of $r$ and $t$ would represent the continuity of (1) current and (2) carrier energy~\cite{kachorovskii2012current,Aizin2016plasmonicboom}. The latter may obtained by integrating the Euler equation across the transient region. However, the length of the transient regions is comparable to the screening length in the 2DEG and is generally smaller than the collision-limited free path. This makes the ballistic description of the transport at the boundary favorable to the hydrodynamic approach. 

Within the ballistic approach, the current across the boundary is calculated as the difference of particle fluxes supplied by the gated and ungated sections. As a result, the variation of current $\delta j$ becomes a linear function of the quasi-Fermi level drop across the boundary, $\delta F_G - \delta F_U$~\cite{Horio-1990-interface_transport}, which should be used as the second boundary condition. The explicit form of this relation is presented in Appendix, Eq.~\ref{eq-rCoefPrecise}. However, the reflection coefficient $r$ can be determined with a sufficient accuracy (see Fig.~\ref{fig-reflection}) if we simply require the continuity of electric potential across the gated/ungated boundary, $\delta \varphi_{\rm g} = \delta \varphi_{\rm u}$. This results in the following expressions for the reflection and transmission coefficients:
\begin{align}
    r_+ = -\frac{1-2\,k_+^g d}{1+2\,k_+^g d}; & \quad r_- = -\frac{k_+^g}{k_-^g}\cdot\frac{1+2\,k_-^g d}{1+2\,k_+^g d}; \label{eq-refl} \\
    t_+ = \frac{4\,k_+^g d}{1+2\,k_+^g d}; & \quad t_-=\frac{1-k_+^g/k_-^g}{1+2\,k_+^g d}.
    \label{eq-transm}
\end{align}

\begin{figure}
    \centering
    \includegraphics[width=\linewidth]{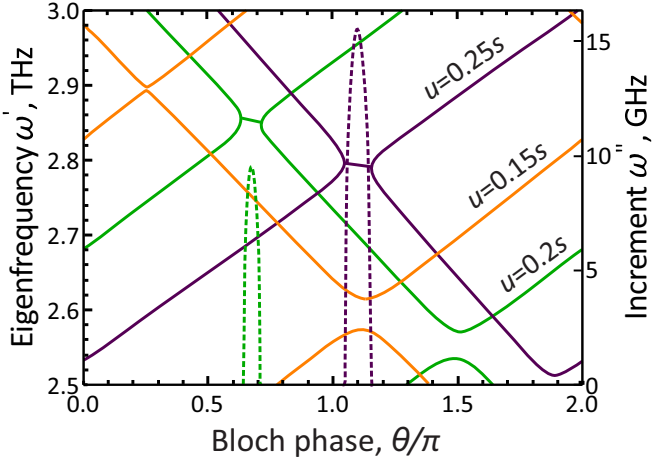}
    \caption{Dispersion curves (solid lines, left scale) and instability increments (dashed lines, right scale) calculated for $\rm GaAs$-based 2DEG with grating gate at different drift velocities. The instability develops from the stable anticrossing (at $u=0.15\,s$) through the merging of the dispersion curves ($u\approx0.17\,s$) to the unstable anticrossing ($u=0.2\,s$). The threshold velocity for instability $u_{th}\approx 0.17\,s$. In the unstable regime, left and right branches of dispersion curve merge through complex frequencies, wherein the real part of frequency is almost independent of quasi-momentum while the imaginary part reaches its maximum at the midpoint. 
    }
    \label{fig-eigenFr_incr}
\end{figure}

With the above reflection and transmission coefficients, the dependence $Z(\omega)$ has a smooth minimum with the minimal value below unity at any non-zero flow velocity, which justifies the neglect of $dZ/d\omega$ in the derivation of the instability condition. As seen from Eqs.~(\ref{eq-Z}), (\ref{ModDepth}) and (\ref{eq-refl}), the threshold velocity for the instability corresponds to the reflectionless passage of the upstream plasmon from the ungated to the gated sections ($r_+ = 0$). At given frequency $\omega$, this velocity is
\begin{equation}
    u_{th}=\left|s-2\,\omega d\right| \approx s\left|1-4\pi\,\frac{d}{\lambda}\right|,
    \label{eq-crVelocity}
\end{equation}
where $\lambda$ is the plasmon wavelength in the gated section. In typical experiments~\cite{knap2004terahertz}, the plasmon wavelength $\lambda$ is on the order of hundreds of nanometers, while the gate-to-channel separation $d$ is several tens of nanometers. Despite tha fact that the ratio $\lambda/d$ is usually small, a large prefactor of $4\pi$ in Eq.~(\ref{eq-crVelocity}) provides an extra order of magnitude to this ratio, and the second term in (\ref{eq-crVelocity}) can even exceed unity.

Fig.~\ref{fig-eigenFr_incr} showing the dispersion curves and plasma instability increments in $\rm GaAs$-based 2DEG under grating gate at different drift velocities substantiates these findings. The lengths of the gated and ungated sections are $L_g=0.6\,\mu\mathrm m$ and $L_u=0.25\,\mu \mathrm m$, respectively, the gate-to-channel separation is $d=10\,\mathrm{nm}$, gate dielectric permittivity $\varepsilon=12.9$, the carrier densities are $n_g=5\cdot 10^{11}\,\mathrm{cm}^{-2}$ and $n_u=2\cdot 10^{12}\,\mathrm{cm}^{-2}$. The onset of the instability represents a transformation of a stable-type plasmon band anticrossing to the unstable one via passing through the gapless plasmon bands. In the unstable case, the neighboring branches of the dispersion curves merge through the complex plane corresponding to two complex conjugate solutions of the dispersion equation and for each Bloch phase. For the parameters of Fig.~\ref{fig-eigenFr_incr}, the unstable mode at 2.9\,THz appears at $u=0.17\,s$ which is in precise agreement with formula (\ref{eq-crVelocity}).


Above the threshold velocity, the waves are unstable for a finite range of the quasi-wave vectors (Bloch phases). Generally, these unstable wave vectors lie away from the edges of the Brillouin zone, i.e. their Bloch phase $\theta = qL \neq 0,\, 2\pi, \, 4\pi$. The reason is that the extrema of the plasmonic bands (being at the edges of Brillouin zone at zero drift velocity) are shifted away by the Doppler effect. Within the unstable domains, the real part of plasmon frequency $\omega'$ almost does not depend on the quasi wave vector, while the increment $\omega''$ varies abruptly. Expanding the dispersion equation (\ref{eq-Dispersion}) near the band anticrossing at $\theta = \theta_{cr}$ and $\omega = \omega_{cr}$, we find
\begin{equation}
    \omega''^2=\frac{2 |\theta-\theta_{cr}|}{(\mathrm d\alpha/\mathrm d \omega)^2}\left|\tan\left(\theta_{cr}+\alpha(\omega_{cr})\right)\right|,
    \label{eq-gamma}
\end{equation}
where $\alpha(\omega)=(k^{\rm g}_+ + k^{\rm g}_-)L_{\rm g}/2+ (k^{\rm u}_+ + k^{\rm u}_-)L_{\rm u}/2$. Equation (\ref{eq-gamma}) describes the square-root growth of increment above the critical Bloch phase observed in Fig.~\ref{fig-eigenFr_incr}.

\begin{figure}
    \centering
    \includegraphics[width=\linewidth]{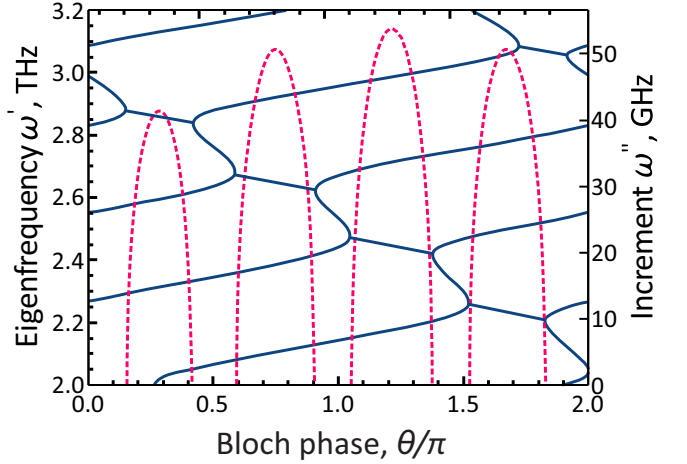}
    \caption{Dispersion curves (solid lines, left scale) and instability increments (dashed lines, right scale) at high drift velocity $u=0.59s$ and other parameters as in Fig.~\ref{fig-eigenFr_incr}.
    }
    \label{fig-denseModes}
\end{figure}

As the drift velocity increases, the branches of plasmon dispersion become denser (Fig.~\ref{fig-denseModes}). This is explained by the fact that the backward wavevector in the gated region grows as $k_-^{\rm g} \propto 1/(u-s)$, and at large velocities even a slight variation of the frequency results in strong variations of the right-hand side of dispersion equation (\ref{eq-Dispersion}). One can also observe that the highest increments are achieved at the frequency, for which the modulation depth $Z$ has the minimal value. This is seen from an arc-shaped envelope of increment curves in Fig.~\ref{fig-denseModes}: the frequency of $\sim 2.4$\,THz corresponds to the minimum of $Z(\omega)$ and to the maximum of the increment. The deviation of frequency from this value leads to an increase in $Z$ and decrease in the increment.

\begin{figure}
    \centering
    \includegraphics[width=\linewidth]{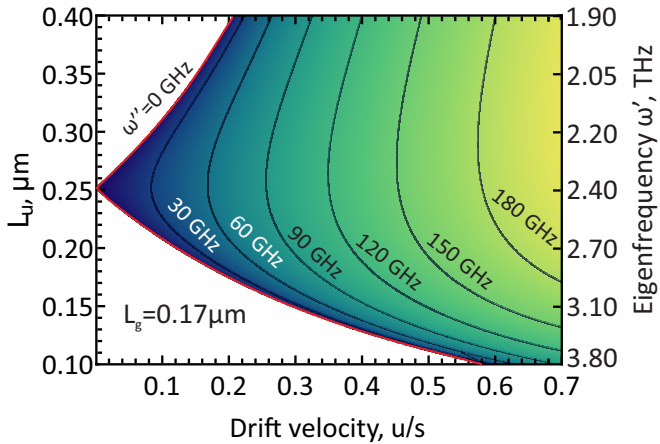}
    \caption{Color map of the instability increment vs drift velocity and the length of the ungated region at the resonant frequency [Eq.~(\ref{eq-omegaRes})]. Gate length $L_{\rm u}=0.17$\,$\mu$m, gate-to-channel separation $d=20$\,nm. The increment at $L_{\rm u}=0.25\,\mu$m is thresholdless (see Eq.~\ref{eq-crVelocity}), and the frequency of the thresholdless instability corresponds to the maximum increment.
    }
    \label{fig-ContourIncr}
\end{figure}

It is remarkable that the maximization conditions for the instability increment can be derived analytically. The maximum instability increment is achieved in the middle of the unstable domain with respect to the quasi momentum $\theta$, and the wider the instability domain is, the larger is the increment. The center of the unstable domain corresponds to zero $\omega$-derivatives of both left- and right-hand sides (lhs. and rhs.) of Eq.~(\ref{eq-Dispersion}). Moreover, the instability increment can be maximized with respect to all other parameters of the problem (gate length, carrier density, etc.) by requiring the maximum difference of lhs. and rhs. of Eq.~(\ref{eq-Dispersion}) at the middle of the unstable domain. In this case, the disbalance between the rhs. and lhs. as functions of real frequencies can be compensated only via introduction of a sufficiently large imaginary part of frequency.

The superposition of these requirements leads us to the following maximum increment conditions
\begin{gather}
    Z\rightarrow \mathrm{min}, 
    \label{eq-boostCondfirst}\\
   \frac{ k^{\rm g}_+ - k^{\rm g}_-}{2}L_{\rm g} = \frac{\pi}{2} +\pi n,\; n\in Z; 
   \label{eq-boostCond1}\\
   \frac{k^{\rm u}_+ - k^{\rm u}_-}{2}L_{\rm u} = \frac{\pi}{2}+\pi m,\; m\in Z. 
    \label{eq-boostCond2}
\end{gather}
The two latter equations can be considered as anti-reflection conditions for coatings represented by gated and ungated regions and, at the same time, the Dyakonov-Shur eigenfrequency conditions for the gated and ungated plasma resonators~\cite{dyakonov1993shallow}. This supports our interpretation of the instability as an amplified DS reflection supplemented by the perfect energy transfer between the cells of plasmonic crystal. 

We further notice that the flow-induced corrections to the phases in Eqs.~(\ref{eq-boostCond1}) and (\ref{eq-boostCond2}) are order of $O(u_{{\rm g},{\rm u}}^2)$. When the carrier density in the ungated region exceeds that in gated region (which corresponds to the range of parameters considered), the phase (\ref{eq-boostCond2}) can be considered as flow-independent. This leads us to the eigenfrequency providing the ultimate increment (at $m=0$)
\begin{equation}
\omega'_{\rm res}=\sqrt{\frac{\pi^2 e^2 n_u}{\varepsilon m^* L_u}}.
\label{eq-omegaRes}
\end{equation}
The peak increment at the resonant frequency is obtained via separating the real and imaginary parts of Eq.~(\ref{eq-Dispersion}) and expanding it with respect to $\omega''$ and $u$. This leads us to
\begin{equation}
\label{OptIncr}
    \omega''^2_{\max}=\frac{2 [1-Z(\omega'_{\rm res})] }{Z(\omega'_{\rm res})T_1^2 + 4 T_1 T_2-(T_1u_1/s)^2},
\end{equation}
where $T_1=s L_g/(s^2-u_g^2)$ and $ T_2=\omega'_{\rm res} L_u/2a$.

The dependence of the ultimate increment on the drift velocity and the length of the ungated region is shown in Fig.~\ref{fig-ContourIncr}. The gate-to-channel separation equals 20\,nm and the gate length is 0.17\,$\mu$m. It can be seen that there exists a length of ungated domain $L^*_u$ and the corresponding resonant frequency $\omega'^*_{\rm res}$ such that the development of instability is thresholdless. Namely, this frequency is $\omega'^*_{\rm res} = s/2d$ (Eq.~ \ref{eq-crVelocity}). At higher drift velocities, the maximum increment is also achieved roughly at that frequency. The regions filled with white in Fig.~\ref{fig-ContourIncr} correspond to the stability of modes with frequency $\omega'_{\rm res}$; the equation of boundary between stable and unstable domains is readily obtained by substituting $\omega = \omega'_{\rm res}$ into the condition $Z(\omega,u)=1$.

\section{Discussion}
For a realistic estimate of the instability increment one has to include damping due to carrier-phonon and carrier-impurity scattering. This can be generally achieved by adding a friction term in Euler equation. For long momentum relaxation rates $\tau_p$ ($\omega\tau_p \gg 1$) the scattering typically reduces the 'collisionless' instability increment by $1/2\tau_p$~\cite{dyakonov1993shallow}. The compensation of collisional damping by the instability increment in GaAs at 77\,K [mobility $\mu = 2\times 10^5$\,cm$^2$/(V s)] occurs at $\omega'' = 65\times 10^9$ s$^{-1}$. Under optimal conditions, this corresponds to the velocity $u^* = 0.25\,s = 1.5\times 10^5$\,m/s. This is twice below the carrier saturation velocity in GaAs at 77\,K.

Another aspect of carrier relaxation is the voltage drop along the channel which may distort the uniform carrier density within a single cell of plasmonic crystal. From the above estimates for GaAs at 77\,K, we find this voltage drop to be $\Delta V \approx u^*L_g / \mu = 1.3$\,mV, which is indeed small compared to the gate voltage required to support the given carrier density under the gates $V_g = ms^2/e = 140$\,mV. Further increase in drift velocity requires considerably larger voltage drops due to the saturation effects. Particularly, $u = 3\times 10^5$ m/s is achieved at $\Delta V\approx 40$ mV. While the density uniformity under a single gate is maintained even in high fields, the uniformity over the whole crystal can be supported by applying a gradually changing voltage to the series of gates with the aid of separative capacitors~\cite{kachorovskii2012current}.

The maximum attainable increment in Fig.~\ref{fig-ContourIncr} of $180\times 10^9$ s$^{-1}$ corresponds to the momentum relaxation time of $2.8$ ps. Such relaxation time characterizes the electron-phonon scattering in graphene at room temperature~\cite{Bolotin2008mobility}. Hence, one might expect the development of amplified-reflection instabilities in graphene-based transistors with grating gates~\cite{Olbrich2016graphenegrating}. We should note, however, that the study of plasma instabilities in graphene requires an essential modification of hydrodynamic equations~\cite{Svintsov2012hydrodynamic,Tomadin2013plasmagraphene} and will be left for future work.

For the quantitative comparison of the presented model and experimental data on THz emission in grating-gate structures~\cite{otsuji2013emission} one needs to consider the reflection of the unstable travelling waves at terminals of plasmonic crystal. The geometrical asymmetry of the plasmonic crystal unit cell (the presence of two gates of unequal length) should be also taken into account. This can be done within the developed transfer matrix formalism, though the resulting dispersion equations are quite cumbersome. The geometrical asymmetry was shown to be crucial for efficient THz detection in plasmonic FETs~\cite{popov2011detectionasymmetric} and is expected to be beneficial to achieve the low-threshold instabilities~\cite{Koseki2016Giant}, though the full theory of the asymmetry effect on the instability has yet to be developed. Further possible extensions of our model include the renouncement of quasi-optical approximation and full electrodynamic treatment of the plasmon reflection at the boundaries, including the excitation of the evanescent waves~\cite{Sydoruk2015reflection}. Within the same formalism, one can also consider the self-excitation of the edge plasmons travelling along the gated/ungated boundary~\cite{Dyakonov2008boundaryinstability}, which might have larger instability increments compared to the bulk modes.

In conclusion, we have theoretically demonstrated the instability of direct current flow in grating-gated 2DES against the excitation of travelling plasmons. The mechanism of instability is associated with the amplified Dyakonov-Shur type plasmon reflection from gated/ungated boundaries and proper phase matching of plasmons under the neighboring gates. Using the transfer matrix approach, we have derived the generic dispersion relation for plasmonic crystals with flow-induced non-reciprocity. In a particular case of alternating gated and ungated regions of 2DEG, this equation has unstable solutions at flow velocities which can be well below the plasma wave and saturation velocities. The increment of predicted instability is order of (but not limited to) 0.15 THz, which makes the instability feasible in GaAs-based 2DEG at 77\,K and in graphene at room temperature.

\section*{Acknowledgement}
The work was supported by the grant No. 16-19-10557 of the Russian Science Foundation. 

\appendix
\section{Ballistic current at the gated/ungated boundary}
Here we will carry out a microscopic calculation of the ballistic current through the gated/ungated (\textit{g/u}) boundary in order to establish the reflection coefficient of an incident plasma wave. We show that plasmon reflection from the boundary can lead to the wave amplification under direct current flow, similar to the reflection from the drain side of the Dyakonov-Shur FET. 

\begin{figure}[hb]
    \centering
    \includegraphics[width=\linewidth]{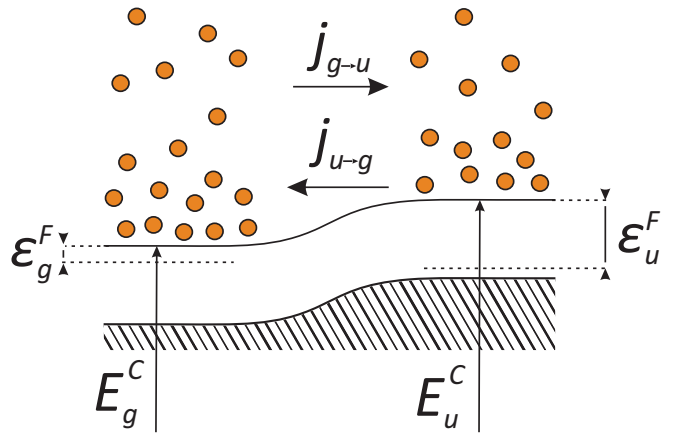}
    \caption{Schematic band diagram of the junction between the gated (left) and ungated (right) regions.
    }
    \label{fig-junction}
\end{figure}

\begin{figure}[hb]
    \centering
    \includegraphics[width=\linewidth]{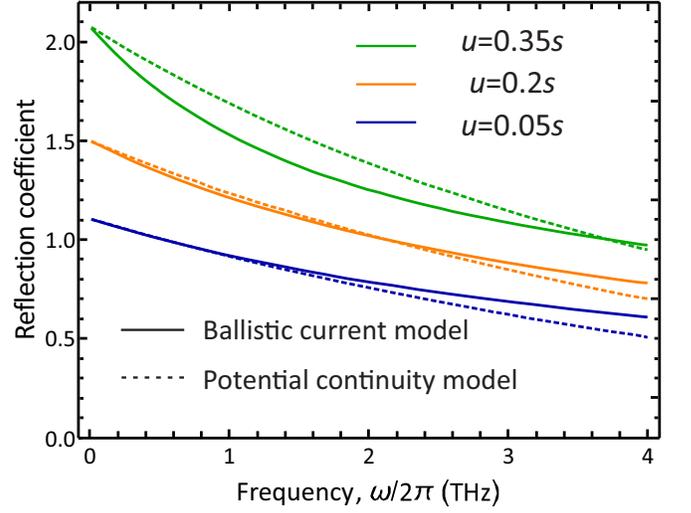}
    \caption{Calculated plasmon reflection coefficient (with respect to the amplitude of potential) from the gated/ungated boundary vs. frequency at different drift velocities in the gated section (in units of plasma wave velocity). Solid lines stand for the reflection coefficients obtained with microscopic calculation of ballistic current at the interface; dashed lines stand for the reflection coefficients obtained by matching of ac potential variations.
    }
    \label{fig-reflection}
\end{figure}

An example of the band diagram of the \textit{g/u} interface is shown in Fig.~\ref{fig-junction}), this corresponds to the enrichment of the gated domain with charge carriers. The net current across the boundary is the difference of carrier fluxes supplied by the gated ($j_{g\rightarrow u}$) and ungated ($j_{u\rightarrow g}$) regions:
\begin{align}
    j_{g\rightarrow u}=\frac{2}{(2\pi\hbar)^2}\int\limits_{p_x>p_{cr}} v_x f(\textbf{p})\mathrm d^2\textbf{p}; \\
    j_{u\rightarrow g}=\frac{2}{(2\pi\hbar)^2}\int\limits_{p_x<0} v_x f(\textbf{p})\mathrm d^2\textbf{p},
\end{align}
where $v$ and $p$ are the electron velocity and momentum, respectively, $p_{cr}=\sqrt{2m(E^c_u-E^c_g)}$ is the minimal momentum required to overcome the barrier at the boundary, $E^c_u$ and $E^c_g$ are the positions of the conduction band bottom in the respective regions. We assume that the carrier distribution obeys the locally equilibrium hydrodynamic function. In accordance with hydrodynamic description, we take the distribution function in the locally equilibrium form 
\begin{equation}
f(\textbf{p})=\exp\left[-\frac{\left({\bf p}-m{\bf u}\right)^2}{2mT}\right],    
\end{equation}
where ${\bf u}$ is the drift velocity and $T$ being the temperature in energy units. Evaluating the integrals, we find the net current $j=j_{g\rightarrow u}+j_{u\rightarrow g}$ across the interface:
\begin{multline}
    j=n_g\left[\frac{e^{-\xi_g^2}}{2\sqrt{\pi}}v_T+\frac{\mathrm{erfc}(\xi_g)}{2}u_g\right]+\\+n_u\left[\frac{e^{-\xi_u^2}}{2\sqrt{\pi}}v_T+\frac{\mathrm{erfc}(\xi_u)}{2}u_u\right].
    \label{eq-currentFull}
\end{multline}
Here $\xi_g=\left(v_{cr}-u_g\right)/v_T$, $\xi_u=u_u/v_T$, $v_T=\sqrt{2T/m}$, $v_{cr}=p_{cr}/m$, $\mathrm{erfc}(x)=2/\sqrt{\pi}\int\limits_{x}^{\infty} e^{-y^2}\,\mathrm dy$ is the complementary error function, and $n_{g,u}$ is the carrier density in the respective region.

Assuming small harmonic perturbations of the quantities $E_g^c$, $E_u^c$, $n_g$, $n_u$, $u_g$, $u_u$ in Eq.~(\ref{eq-currentFull}), we find the microscopic boundary condition relating the ac current across the boundary $\delta j$ to the ac variations of density, velocity, and electric potential:
\begin{widetext}
\begin{multline}
\label{Full-J}
    \delta j= \delta n_g\left[\frac{e^{-\xi_g^2}}{2\sqrt{\pi}}v_T + \frac{\mathrm{erfc}(\xi_g)}{2}u_g\right] + \delta n_u\left[\frac{e^{-\xi_u^2}}{2\sqrt{\pi}}v_T + \frac{\mathrm{erfc}(\xi_u)}{2}u_u\right] + \\ + n_g v_T\cdot\frac{e^{-\xi_g^2}}{2\sqrt{\pi}}\frac{\left(\delta E^C_g - E^C_u\right)}{T} + n_g \delta u_g\left[\frac{1}{\sqrt{\pi}}\frac{v_{cr}}{v_T}e^{-\xi_g^2} + \frac{\mathrm{erfc}(\xi_g)}{2}\right] + n_u\delta u_u\cdot\frac{\mathrm{erfc}(\xi_u)}{2}.
\end{multline}
\end{widetext}
The small-signal variations of density in (\ref{Full-J}) can be expressed through the variations of electric potential with Poisson's equation. Finally, Eq.~(\ref{Full-J}) and supplemented by the continuity of current allows one to obtain the reflection coefficient for plasmon propagating along the direction of drift from the gated to ungated boundary:
\begin{equation}
    r=-\frac{\alpha_+-\beta/(2\,k_+^gd)}{\alpha_- -\beta/(2\,k_-^g d)}, 
    \label{eq-rCoefPrecise}
\end{equation}
where
\begin{multline}
    \alpha_\pm=\left(s^2\mp sv_{cr} +v_T^2/2\right)e^{-\xi_g^2} - \\ -\sqrt{\pi} v_T(s\pm u_g) + (u_g\mp s)v_T \frac{\sqrt{\pi}\mathrm{erfc}(\xi_g)}{2};
\end{multline}
\begin{multline}
    \beta = v_T^2\cdot k_+^u de^{-\xi_u^2} + s^2e^{-\xi_g^2} + \frac{\sqrt{\pi}}{2}\mathrm{erfc}\left(\xi_u\right)v_T u_u\times \\ \times \left[-2k_+^u d + \frac{s^2}{u_g^2}\left(\frac{n_u}{n_g}\right)^3\frac{k_+^u u_u}{\omega-k_+^u u_u}\right].
\end{multline}
The reflection coefficients calculated with Eq.~(\ref{eq-rCoefPrecise}) are shown in Fig.~\ref{fig-reflection} with solid lines. Instead of using the cumbersome microscopic condition (\ref{Full-J}), one can require the continuity of quasi-Fermi level across the boundary which is commonly used in the modelling of transport across the heterojunctions~\cite{Horio-1990-interface_transport}. Moreover, the variations of carrier Fermi energy are typically small compared to the variations of electric potential, $\delta\varepsilon_F/e\delta\varphi \approx v_T/s \ll 1$. In such situation, the continuity of the quasi-Fermi level implies the continuity of the electric potential. The reflection coefficient for the potential continuity boundary condition is just Eq.~(\ref{eq-rCoefPrecise}) with $\alpha_+ = \alpha_- = \beta = 1$.

The comparison of reflection coefficients calculated with Eq.~(\ref{eq-rCoefPrecise}) and with simplified model of potential continuity is presented in Fig.~\ref{fig-reflection}. The discrepancy between these two results is less than $10$\%, hence, the continuity of electric potential can be used as a boundary condition with sufficient accuracy.

\bibliography{Petrov_Autumn_2016}

\end{document}